\documentclass{nature}
\usepackage{graphicx}
\makeatletter
\let\saved@includegraphics\includegraphics
\AtBeginDocument{\let\includegraphics\saved@includegraphics}
\renewenvironment*{figure}{\@float{figure}}{\end@float}
\makeatother
\usepackage{bm}% bold math
\usepackage{ulem}
\usepackage{xcolor}

\title{Colossal electromagnon excitation in the non-cycloidal phase of TbMnO$_3$ under pressure}

\begin{document}

\author{Ian Aupiais$^1$, Masahito Mochizuki$^2$, Hideaki Sakata$^3$, Romain Grasset$^{1,4}$, Yann Gallais$^1$, Alain Sacuto$^1$ \& Maximilien Cazayous$^1$}

\maketitle

\begin{affiliations}
\item Laboratoire Mat\'eriaux et Ph\'enom\`enes Quantiques (UMR 7162
 CNRS), Universit\'e Paris Diderot-Paris 7, 75205 Paris cedex 13, France
\item Department of Applied Physics, Waseda University, Shinjuku-ku Tokyo 169-8555, Japan
\item Department of Physics, Tokyo University of Science, 1-3 Kagurazaka Shinjyuku-ku Tokyo 162-8601, Japan
\item Present address: Laboratoire des Solides Irradi\'es, Ecole polytechnique, Route de Saclay 91128 Palaiseau, France
\end{affiliations}

\section*{ABSTRACT}
\begin{abstract}
The magnetoelectric coupling, i.e., cross-correlation between electric and magnetic orders, is a very desirable property to combine functionalities of materials for next-generation switchable devices. Multiferroics with spin-driven ferroelectricity presents such a mutual interaction concomitant with magneto- and electro-active excitations called electromagnons. TbMnO$_3$ is a paradigmatic material in which two electromagnons have been observed in the cycloidal magnetic phase. However, their observation in TbMnO$_3$ is restricted to the cycloidal spin phase and magnetic ground states that can support the electromagnon excitation are still under debate. 
Here, we show by performing Raman spectroscopy measurements under pressure that the lower-energy electromagnon (4 meV) disappears when the ground state enters from a cycloidal phase to an antiferromagnetic phase (E-type). On the contrary, the magneto-electric activity of the higher-energy electromagnon (8 meV) increases in intensity by one order of magnitude. Using microscopic model calculations, we demonstrate that the lower-energy electromagnon, observed in the cycloidal phase, originates from a higher harmonic of the magnetic cycloid, and we determine that the symmetric exchange-striction mechanism is at the origin of the higher-energy electromagnon which survives even in the E-type phase. The colossal enhancement of the electromagnon activity in TbMnO$_3$ paves the way to use multiferroics more efficiently for generation, conversion and control of spin waves in magnonic devices.
\end{abstract}
 
\maketitle
\section*{INTRODUCTION}
In the so-called improper multiferroics such as perovskite manganites RMnO$_3$ with R being a rare-earth ion, the electric polarization is induced by a spin order breaking the spatial inversion symmetry via the spin-orbit interaction \cite{Mostovoy2006, Sergienko2006a,  Mochizuki2009, Kenzelmann2005, Hu2008} or the magnetic exchange striction \cite{Sergienko2006b, Picozzi2007}. TbMnO$_3$ is one of the most studied multiferroic materials of this class. Such a compound is fundamental to study novel coupling between microscopic degrees of freedom such as spin and charge \cite{Khomskii2009}. In 2006, Pimenov \textit{et al.} \cite{Pimenov2006} succeeded in exciting spin waves with the electric-field component of terahertz (THz) light in TbMnO$_3$. They called these excitations electromagnons, excitations theoretically put forth by Smolenskii and Chupis \cite{Smolenskii1982} more than 20 years ago. They also showed that the electromagnons could be suppressed by applying a magnetic field, directly demonstrating magnetic-field-tuned electric excitations \cite{Pimenov2006, Pimenov2009}. Their signatures have been evidenced by a large number of techniques : THz \cite{Pimenov2006, Pimenov2009}, infrared (IR) \cite{Aguilar2009, Takahashi2008}, Raman \cite{Rovillain2010} spectroscopies and inelastic neutron scattering \cite{Senff2007}. The interest for the electromagnons extends to applications. Tuning magnetic properties of multiferroics is the first step to build a new technology based on spin waves called magnonics using these promising materials \cite{Kruglyak2010, Eerenstein2006}. In particular, it has been demonstrated that the electromagnons can be exploited to directly manipulate atomic-scale magnetic structures in the rare-earth manganites using THz optical pulses \cite{Kubacka2014}.

Considerable efforts have been devoted to determine the origin of the electromagnons. Katsura \textit{et al.} \cite{Katsura2007} firstly proposed a model for a dynamical coupling based on the inverse Dzyaloshinskii-Moriya (DM) interaction, originally developed to explain the static ferroelectricity in TbMnO$_3$ \cite{Sergienko2006a}. However, the light polarization predicted to observe such excitations is in contradiction with the two electromagnons observed in IR \cite{Aguilar2009} and THz spectroscopies \cite{Pimenov2009}. The two electromagnons are observed around 60 cm$^{-1}$ (7.4 meV) and 35 cm$^{-1}$ (3.7 meV) in the cycloidal magnetic phase. The higher-energy electromagnon has already been explained as a zone-edge magnon activated purely by the magnetostriction mechanism \cite{Aguilar2009, Stenberg2009, Mochizuki2010a}, but the origin of the lower-energy electromagnon is still under debate. Two models have been proposed, one based on the anharmonic component of the cycloidal ground state \cite{Mochizuki2010a}, and another assuming an anisotropic magnetostriction coupling \cite{Stenberg2009}.
\\
\\
In this work, we investigate the dynamical part of the magnetism in TbMnO$_3$ under hydrostatic pressure to determine the exact spin ground state responsible for the electromagnon activity. At ambient pressure and below $T_{\rm C} = 28$K, the Mn magnetic moments exhibits an incommensurate cycloidal magnetic order propagating along the $b$ direction with a wave vector $Q_{\rm C}=(0,0.28,1)$ \cite{Kenzelmann2005}. The magnetic order combined with an inverse DM interaction induces an electrical polarization along the $c$ axis \cite{Sergienko2006b, Kimura2003, Walker2011}. At around 5 GPa and 10K, the spin ground state changes from this cycloidal state to the E-type antiferromagnetic state with a wave vector $Q_{\rm E}=(0,0.5,1)$ \cite{Makarova2011, Terada2016}, in which a giant spin-driven ferroelectric polarization has been observed along the $a$ axis \cite{Aoyama2014, Aoyama2015}. Tracking the electromagnons, low energy excitations with small Raman intensity, requires the development of dedicated instrumentation to optically study them under pressure. The experimental setup is based on a diamond anvil cell in a non-colinear scattering geometry. It allows to increase the numerical aperture collection and reduces the background signal to measure small Raman signal at low energy. We have then been able to measure the electromagnon Raman signal on TbMnO$_3$ as low as 10 cm$^{-1}$ and up to 8 GPa. 

\section*{RESULTS}
\subsection{Experiments} \hfill

\begin{figure}[tb!]
\begin{center}
\includegraphics[width=13cm]{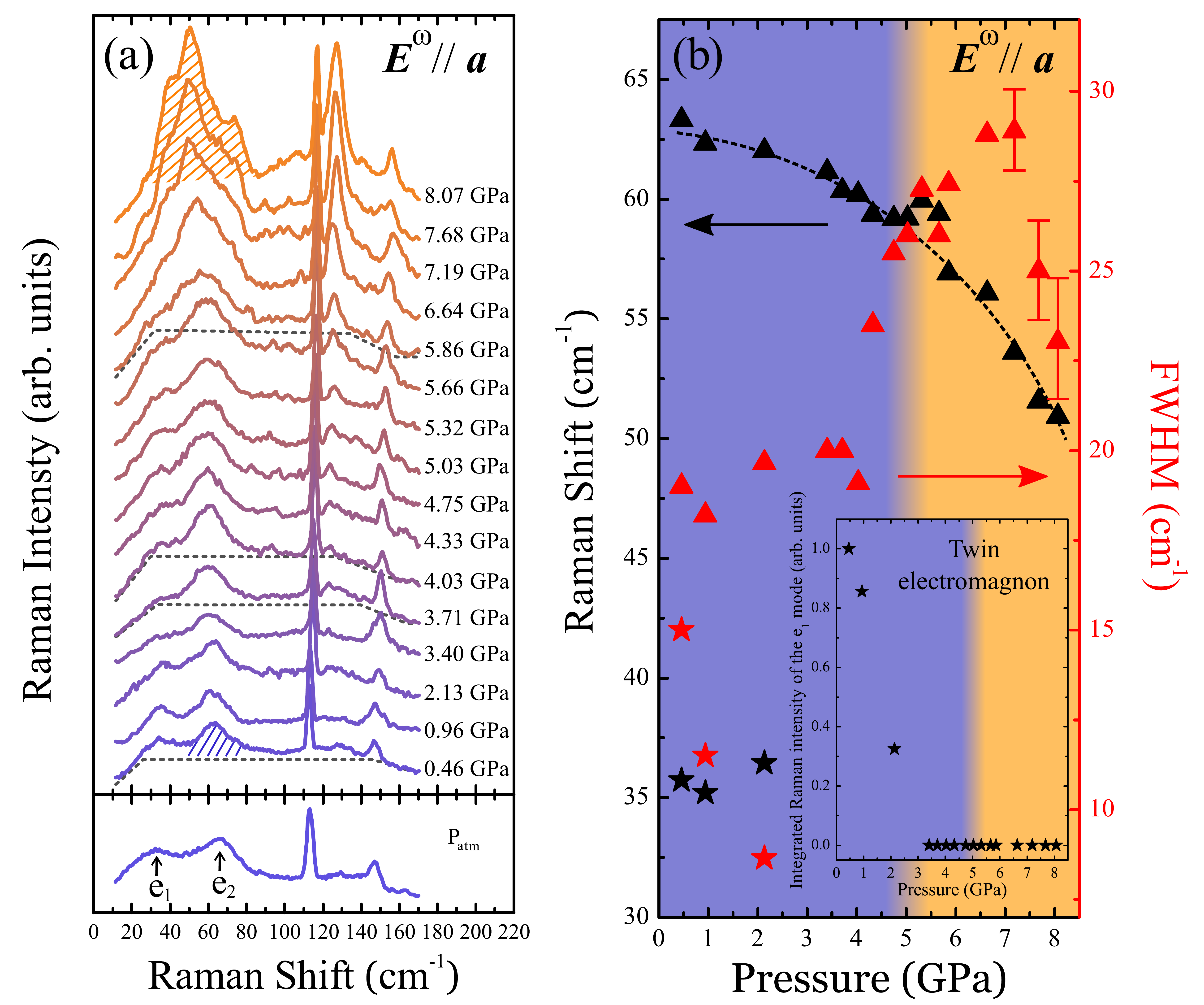}
\caption{\textbf{Raman signature of electromagnons along the $a$ axis in TbMnO$_3$ under hydrostatic pressure.} (\textbf{a}) Low-energy Raman spectra at 11K for different hydrostatic pressures for A$_{\rm g}$ symmetry with a light polarization along the $a$ axis. The spectrum at 0 GPa has been measured outside the diamond anvil cell. The e$_1$ electromagnon disappears between 2.13 GPa and 3.40 GPa (see Supplementary Information). Dashed lines correspond to the backgrounds. (\textbf{b}) Raman shift and Full Width of Half Maximum (FWHM) of the e$_1$ (stars) and the e$_2$ (triangles) electromagnons. Dashed line is a guide for the eyes. The insert is a plot of the integrated Raman intensity of the e$_1$ electromagnon as a function of pressure.}
\label{fig:1}
\end{center}
\end{figure}

Figure \ref{fig:1}a shows the low-energy part of the A$_{\rm g}$ Raman spectra at 11 K for different pressures with light polarizations parallel to the $a$ axis (for more details see Methods). At 0 GPa, two low energy excitations, associated with electromagnons, are observed at 60 cm$^{-1}$ (e$_2$) and 35 cm$^{-1}$ (e$_1$). The mode at 60 cm$^{-1}$ corresponds to the zone-edge magnon of the cycloid activated by the pure exchange-striction mechanism \cite{Aguilar2009,Mochizuki2010a}. Even if its origin is still under debate, the mode at 35 cm$^{-1}$ has been attributed to a magnon located at $2Q_{\rm C}$ away from the zone-edge, which corresponds to a replica of the e$_2$ mode and hence is referred to as a twin electromagnon \cite{Stenberg2009,Mochizuki2010a}. As seen in the insert of Fig. \ref{fig:1}b, the integrated Raman intensity of the twin electromagnon (e$_1$) decreases as the pressure increases and finally it disappears between 2.13 GPa and 3.4 GPa. On the contrary, the intensity of the zone-edge electromagnon (e$_2$) is strongly enhanced above 4.5 GPa. Its energy decreases continuously as the pressure increases. The FWHM of the e$_2$ electromagnon in Fig. \ref{fig:1}b presents a step around 4.5 GPa, corresponding to the transition from the cycloidal phase to the E-type phase. The transition between the cycloid and the E-type phase is expected to be sharp. However, we suspect the transition is smoothed by domains which are averaged under our laser spot (size of 30 $\rm \mu m$). Peaks above 100 cm$^{-1}$ are ascribed to phonon modes. The frequencies of the two phonons at 110 cm$^{-1}$ and 145 cm$^{-1}$ increase under pressure due to the decrease of the crystal volume. We notice that an additional peak appears at 124 cm$^{-1}$ above 4.75 GPa. Since no structural transition has been reported at this pressure previously, the activation of this mode is not trivial. This might be a Raman-activated infrared phonon mode coupled to magnetic excitations or a two-magnon excitation coupled with phonon modes \cite{Takahashi2008}.

\begin{figure}[tb!]
\begin{center}
\includegraphics[width=8cm]{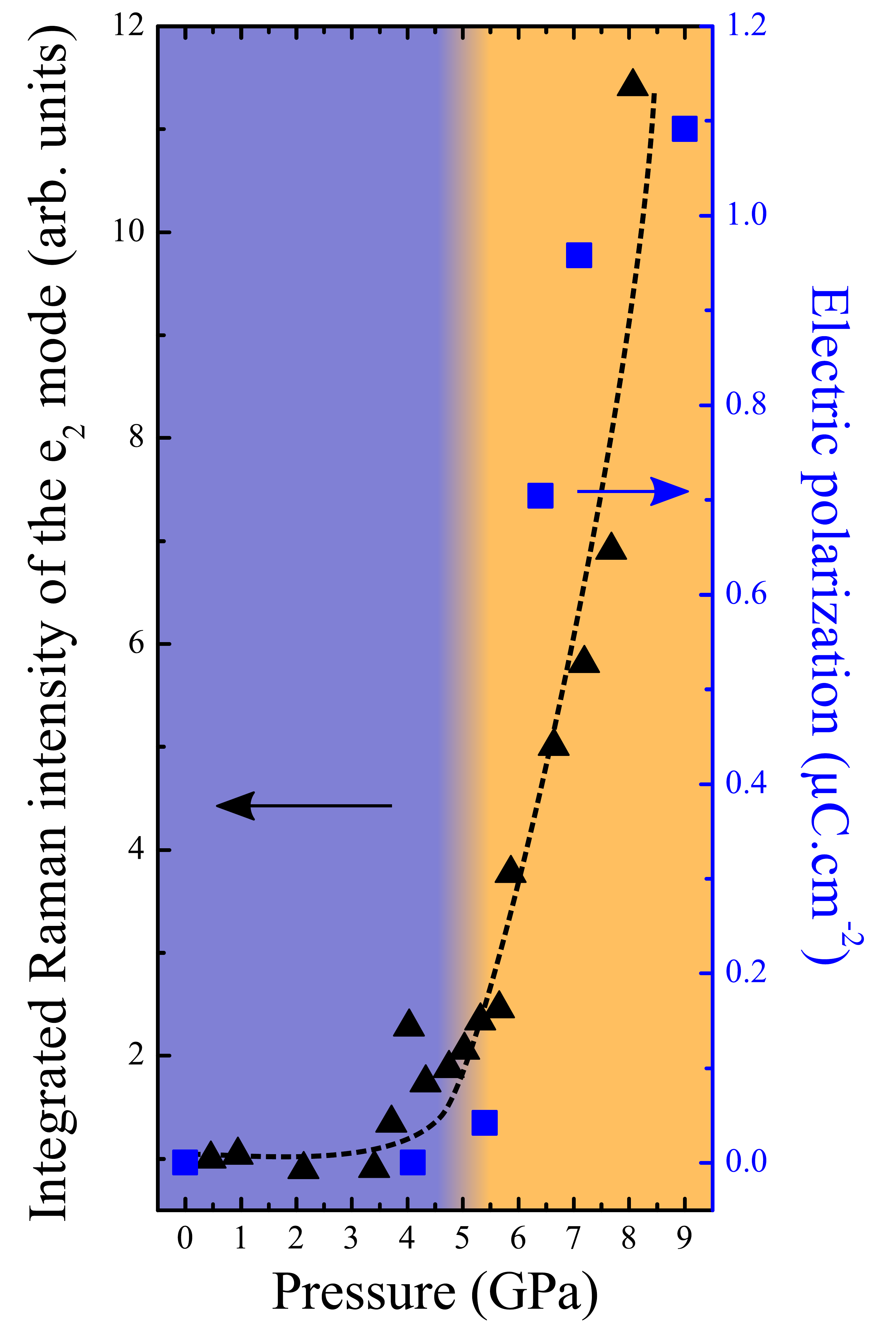}
\caption{\textbf{Colossal enhancement of the electromagnon activity in the E-type phase of TbMnO$_3$.} Integrated Raman intensity of the zone-edge e$_2$ electromagnon (triangles) as a function of the pressure. Measured spontaneous ferroelectric polarization (squares) in the a direction as functions of the pressure at 10 K extracted from T. Aoyama \textit{et al.} \cite{Aoyama2015}. Dashed line is a guide for the eyes.}
\label{fig:2}
\end{center}
\end{figure}

Figure \ref{fig:2} represents the integrated Raman intensity of the zone-edge e$_2$ electromagnon peak extracted from Fig. \ref{fig:1}a as a function of the pressure. The integrated intensity of the e$_2$ mode is a signature of the electro- and magneto-activity of the electromagnon. Above the transition to the E-type phase at 4.5 GPa, the activity of this electromagnon increases and is multiplied by an order of magnitude at 8.07 GPa. Moreover, a jump is observed around 4.5 GPa which reflects the switching of the ferroelectric polarization vector from the $c$ axis to the $a$ axis as seen by the P-E ferroelectric hysteresis loop\cite{Aoyama2014, Aoyama2015}. The measured spontaneous ferroelectric polarization\cite{Aoyama2015} under pressure is included in Fig. \ref{fig:2}. We find that both set of data follow a similar behavior. The experimental polarization reaches a maximum value of 1.1 $\rm \mu$C.cm$^{-2}$ at 9 GPa, an order of magnitude larger than the value at ambient pressure. It is clear that the continuous increase of the zone-edge electromagnon activity is correlated with the emergence and the hardening of the electric polarization along the $a$ axis observed in the E-type phase. This underlines that the mechanism of the spin-driven ferroelectricity is also involved in the origin of the electromagnon activity in this phase. 

\subsection{Theory} \hfill

\begin{figure}[tb!]
\begin{center}
\includegraphics[width=12cm]{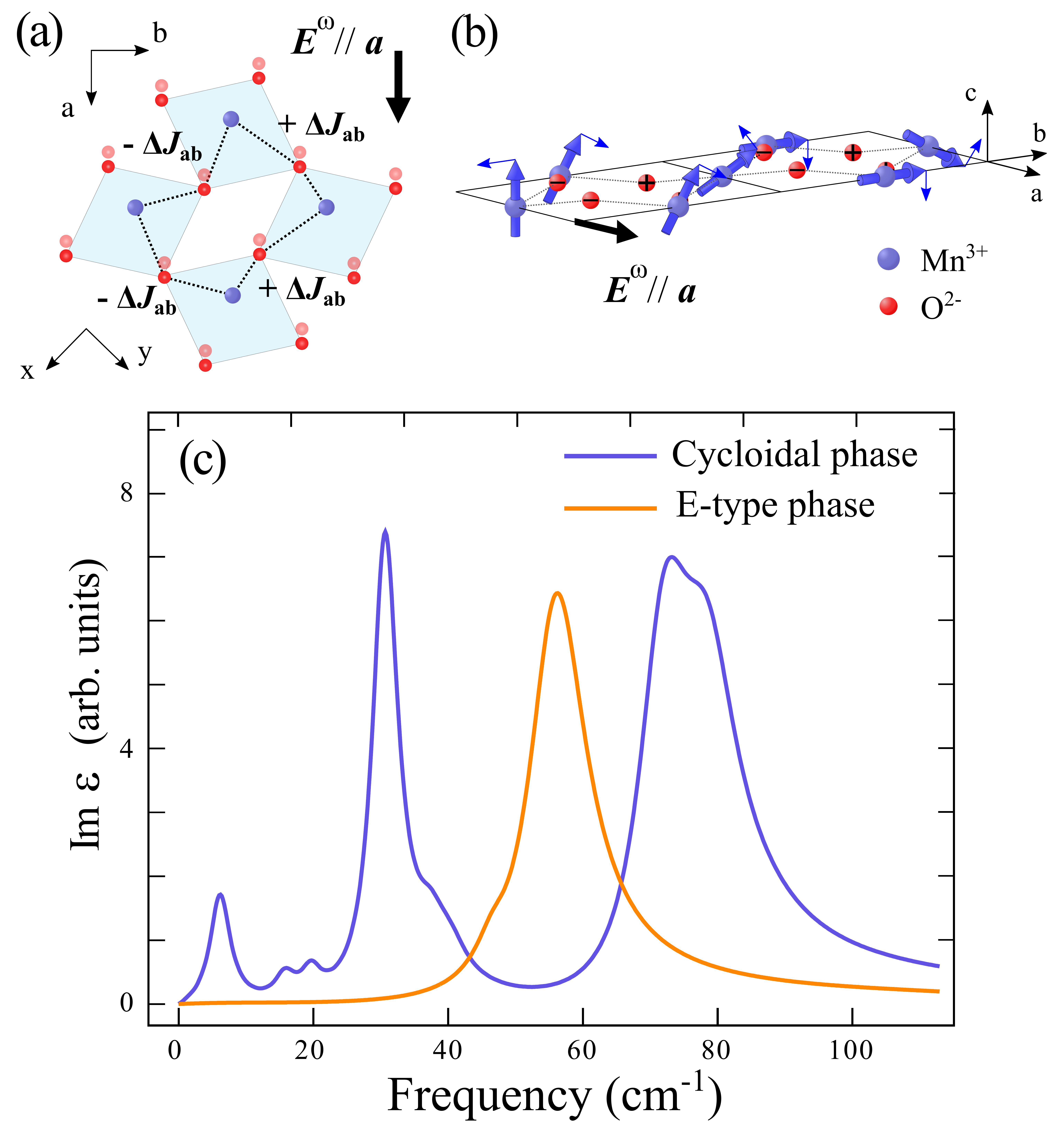}
\caption{\textbf{Theoretical calculations based on the microscopic spin model.} (\textbf{a}) Coupling between the electric field of light, $\bm E^{\rm \omega} \parallel \bm a$, and magnetic order in the $ab$ plane via the symmetric exchange-striction mechanism. (\textbf{b}) The zone-edge magnon induced by this mechanism, small blue arrows represent the out of phase oscillation of the spins. (\textbf{c}) Electromagnon absorption spectra calculated for the $bc$-plane cycloidal phase under a low pressure and the E-type antiferromagnetic phase under a high pressure.}
\label{fig:3}
\end{center}
\end{figure}

To shed light on a physical mechanism of the electromagnons in the E-type phase, we developed calculations based on microscopic spin Hamiltonian consisting of five termes as $H=H_{\rm ex}+H_{\rm DM}+H_{\rm sia}^{\rm D}+H_{\rm sia}^{\rm E}+H_{\rm biq}$. The first term $H_{\rm ex}$ describes superexchange interactions for which four kinds of exchange couplings are taken into account. The competition between the ferromagnetic coupling $J_{\rm ab}$ and the antiferromagnetic coupling $J_{\rm b}$ in the $ab$ plane induces a cycloidal magnetic order. The interplane antiferromagnetic coupling $J_{\rm c}$ causes a staggered stacking along the $c$-axis. A ferromagnetic coupling J$_{\rm a}$ along the $a$-axis is neccessary to reproduce the E-type phase although it plays only a minor role in the cycloidal phase. $H_{\rm DM}$ is the DM interaction. The terms $H_{\rm sia}^{\rm D}$ and $H_{\rm sia}^{\rm E}$ describe the single-ion magnetic anisotropies, both of which introduces anharmonicity to the magnetic cycloid. The last  term $H_{\rm biq}$ represents the biquadratic interaction, which reinforces anharmonicty in the cycloidal phase and stabilizes the E-type phase (see Supplementary Information).
\\
\\
Spin- and helicity-correlation functions are calculated to identify the magnetic order. The low-pressure phase is generated with $J_{\rm b}=0.64$ meV which reporduces the $bc$-plane cycloid with anharmonic components. Application of pressures to TbMnO$_3$ decreases the crystal volume and reduces the bond length \cite{Aoyama2014}. The exchange paths of $J_{\rm ab}$ and $J_{\rm b}$ are Mn(1)-O-Mn(2) and Mn(1)-O-O-Mn(3), respectively (see Supplementary Information). Hence the former is mediated by one O ion, whereas the latter by two O ions with p-p hybridization between their 2p orbitals. Because the p-p hybridization increases more siginificantly with pressure than the d-p hybridization, we simulate the effect of pressure on the magnetic structure by increasing the value of $J_{\rm b}$. The electric polarization in the E-type phase results from the oxygen displacements in the $ab$-plane along the $x$ or $y$ axis due to symmetric exchange striction, the so-called symmetric exchange-striction mechanism. Therefore we adopt a bond alternation for the in-plane Mn-O-Mn bonds as $J_{\rm ab} \pm \Delta J_{\rm ab}$. To generate the high-pressure phase we set $J_{\rm b}=1.20$ meV and $\Delta J_{\rm ab}=0.04$ meV, we obtain the commensurate E-type antiferromagnetic phase with spins pointing in the $b$ direction.
\\
\\
The electromagnon and magnon excitations has been calculated with the microscopic spin model. We consider the coupling $-{\bm E}^{\rm \omega} \cdot \bm P$ between the external electric field ${\bm E}^{\rm \omega}$ and the spin-dependent electric polarization $\bm P$ given by:
\begin{equation}
P_{\rm \gamma}=\Pi_{\rm \gamma} \sum\limits_{i} [(-1)^{i_x+i_y+m}\bm S_i \cdot \bm S_{i+x}+(-1)^{i_x+i_y+n}\bm S_i \cdot \bm S_{i+y}]
\label{Eq:1}
\end{equation}
where $(m,n)=(0,0)$ for $\gamma=a$, $(m,n)=(1.0)$ for $\gamma=b$, $(m,n)=(i_z+1,i_z+1)$ for $\gamma=c$.
\\
\\ 
In the E-type phase, this term shifts oxygen atoms along the $x$ and $y$ axes and gives rise to a ferroelectric polarization along the $a$ axis as observed experimentally~\cite{Aoyama2014, Aoyama2015}. We first prepare spin configurations at a low temperature by the Monte-Carlo thermalization, and then let the spins configuration relaxed with a sufficient time evolution using the Landau-Lifshitz-Gilbert (LLG) equation. We apply a short pulse of electric field ${\bm E}^{\omega} \parallel \bm a$ to the relaxed system at $t$=0 and trace the time evolution of ferroelectric polarization by numerically solving the LLG equation using the fourth-order Runge-Kutta method (see Supplementary Information). The Fourier transform of the obtained time profiles of the spin-dependent electric polarization gives the electromagnon absorption spectrum. 
\section*{DISCUSSION}
The calculated electromagnon spectra for the two phases are displayed in Fig. \ref{fig:3}c. We obtain two modes at 30 cm$^{-1}$ and 80 cm$^{-1}$ in the $bc$-plane cycloidal phase. These two modes correspond to the twin (e$_1$) and the zone-edge (e$_2$) electromagnons, respectively. On the other hand, we obtain only one mode at 60 cm$^{-1}$ in the E-type phase which corresponds to the down shifted e$_2$ electromagnon. Aguilar \textit{et al.} \cite{Aguilar2009} argued that the exchange-striction mechanism does not work in the collinear spin phase. Since the E-type phase has long been believed to be a collinear order, we expect that the electromagnon excitation mediated by the exchange-striction mechanism should disappear in the E-type phase. However, our calculation reproduced a large electromagnon resonance in the E-type phase. This can be understood from the non collinear nature of the E-type order. The Mn spins in the E-type phase are not perfectly collinear but are considerably canted to form a depressed commensurate ab-cycloid as found in previous theoretical studies \cite{Mochizuki2011, Mochizuki2010b}. Importantly, this canted E-type phase has a magnetic periodicity of $Q_{\rm E}=(0,0.5,1)$ that fits with a crystal periodicity of the orthorhombically distorted perovskite lattice of TbMnO$_3$. Therefore, the E-type order does not contain any higher harmonic components, which explains the absence of the lower-lying mode observed in the incommensurate cycloidal phase because it originates from the Brillouin-zone folding due to the magnetic higher harmonics of the cycloidal order \cite{Mochizuki2010a}. As a result, the electromagnon spectrum has a single peak corresponding to the higher-lying zone-edge mode only.
\\
\\
These results are in good agreement with our experimental data shown in Fig. \ref{fig:1}. In the E-type phase, the e$_1$ electromagnon disappears, whereas the e$_2$ electromagnon has its frequency downshifted. This evidences that the anharmonicity is a mandatory condition for emergence of the twin electromagnon. The fact that the twin electromagnon (e$_1$) disappears before the transition may be due to weakening of the anarhominicity of the cycloid or/and to the spatial coexistence of the cycloid and the E-type state in the vicinity.
\\
\\
In conclusion, we investigated the dynamical magnetoelectric properties of TbMnO$_3$ under hydrostatic pressure with both Raman spectroscopy and microscopic model calculations. We find that the lower-lying e$_1$ electromagnon in the anharmonic cycloidal order disappears in the E-type phase for which anharmonic components are absent. Our finding provides the evidence that the activation of the low-energy electromagnon requires an anharmonicity of the cycloid in TbMnO$_3$. As in the case of the multiferroic BiFeO$_3$\cite{Desousa2008, Fishmann2013}, the anharmonicity is the key to understand the finest properties of the cycloidal multiferroics. We also have shown that an electrical polarization induced by the exchange-striction mechanism increases the activity of the zone-edge electromagnon by one order of magnitude. Such conditions have been realized at ambient pressure in strained TbMnO$_3$ thin films \cite{Shimamoto2017} in which enhanced electromagnon excitations might be observed, providing more efficient building blocks for magnonics devices.

\section*{METHODS}
\subsection{Samples} \hfill

Single crystals of TbMnO$_3$ were grown by floating-zone method and aligned using Laue X-ray back-reflection. The crystals have been polished to obtain high surface quality for optical measurements. TbMnO$_3$ crystallizes in the orthorhombic symmetry (P\textit{bnm}) with lattice parameters equal to $a=5.3~ \rm \mathring{A}, b=5.86~ \rm \mathring{A}, c=7.49~\rm \mathring{A} $.\cite{Alonso2000} 
TbMnO$_3$ becomes antiferromagnetic below the N\'eel temperature $T_{\rm N}=42K$ \cite{Quezel1977}. In this phase the Mn magnetic moments form an incommensurate sinusoidal wave with a modulation vector along the $b$ axis. The ferroelectric order appears below $T_{\rm C}=28K$ where the Mn magnetic moment transit to an incommensurate cycloidal phase \cite{Kenzelmann2005, Kimura2005}. In this phase the spin of Mn$^{3+}$ rotates in the $bc$ plane, and the ferroelectric polarization appears along the $c$ axis. We have probed one TbMnO$_3$ single crystals with a $ac$ plane.
\subsection{Light scattering} \hfill

Raman scattering measurements are performed in a diamond anvils cell equipped with a membrane for change of the hydrostatic pressure. The fluorescence of ruby balls is used as a pressure gauge. The pressure transmitting medium is helium. The incident laser spot is about 20 $\mu m$ diameter size. We have used a triple spectrometer Jobin Yvon T64000 equipped with a liquid-nitrogen-cooled CCD detector and solid laser with a line at 561~nm. 
\section*{DATA AVAILABILITY}
The authors declare that [the/all other] data supporting the findings of this study are available within the paper [and its supplementary information files].
\section*{ACKNOWLEDGEMENTS}
M. Mochizuki thanks supports from JSPS KAKENHI (Grant No. 17H02924), Waseda University Grant for Special Re-search Projects (Project Nos. 2017S-101, 2018K-257), and JST PRESTO (Grant No. JPMJPR132A). The authors want to thank Marie Aude Measson for her technical support. 
\section*{AUTHOR CONTRIBUTIONS}
The samples were grown by H.S.; I.A., R.G. and M.C. performed experiments; I.A. and M.C. analyzed data; M.M developed the model and analyzed data. All authors discussed the results and wrote the manuscript.
\subsection{Competing Interests:} the Authors declare no Competing Financial or Non-Financial Interests. 

\subsection{Correspondence:} Correspondence and requests for materials should be addressed to M. Cazayous~(email: maximilien.cazayous@univ-paris-diderot.fr).
\\
\\

\end{document}